\documentclass[aps,prl,twocolumn,superscriptaddress,amssymb,showpacs]{revtex4}

\usepackage{graphicx}

\begin{document}

% Use the \preprint command to place your local institutional report
% number in the upper righthand corner of the title page in preprint mode.
% Multiple \preprint commands are allowed.
% Use the 'preprintnumbers' class option to override journal defaults
% to display numbers if necessary
%\preprint{}

%Title of paper
\title{Strong continuous-variable entanglement in modulated quantum dissipative dynamics}

% repeat the \author .. \affiliation  etc. as needed
% \email, \thanks, \homepage, \altaffiliation all apply to the current
% author. Explanatory text should go in the []'s, actual e-mail
% address or url should go in the {}'s for \email and \homepage.
% Please use the appropriate macro foreach each type of information

% \affiliation command applies to all authors since the last
% \affiliation command. The \affiliation command should follow the
% other information
% \affiliation can be followed by \email, \homepage, \thanks as well.

\author{H.~H.~Adamyan}
\email[]{adam@unicad.am}
%\homepage[]{Your web page}
%\thanks{}
%\altaffiliation{}
\affiliation{Yerevan State University, A. Manookyan 1, 375049,
Yerevan, Armenia} \affiliation{Institute for Physical Research,
National Academy of Sciences,\\Ashtarak-2, 378410, Armenia}

\author{G.~Yu.~Kryuchkyan}
\email[]{gkryuchk@server.physdep.r.am}
%\homepage[]{Your web page}
%\thanks{}
%\altaffiliation{}
\affiliation{Yerevan State University, A. Manookyan 1, 375049,
Yerevan, Armenia} \affiliation{Institute for Physical Research,
National Academy of Sciences,\\Ashtarak-2, 378410, Armenia}

%Collaboration name if desired (requires use of superscriptaddress
%option in \documentclass). \noaffiliation is required (may also be
%used with the \author command).
%\collaboration can be followed by \email, \homepage, \thanks as well.
%\collaboration{}
%\noaffiliation

% insert suggested PACS numbers in braces on next line
% insert suggested keywords - APS authors don't need to do this
%\keywords{}

\begin{abstract}
We investigate the creation of entangled states of bright light
beams obeying the condition of strong Einstein-Podolsky-Rosen-like
paradox criterion in time-modulated quantum dissipative dynamics.
Having in view the generation of these states we propose a
non-degenerate optical parametric oscillator(NOPO) driven by an
amplitude-modulated pump field. We develop semi-classical and
quantum theories of this device for all operational regimes
concluding that, contrary to ordinary NOPO, the
continuous-variable entanglement becomes approximately perfect for
high level of modulation. Our analytical results are in well
agreement with numerical simulations.
\end{abstract}

%\maketitle must follow title, authors, abstract, \pacs, and \keywords
\maketitle

Continuous-variable (CV) entangled states of light beams provide
excellent tools for testing the foundations of quantum physics and
arouse growing interest due to apparent usefulness as a promising
technology in quantum information and communication protocols
\cite{Braunstein, Furusawa}. The efficiency of quantum information
schemes and quantum measurements significantly depends on the
degree of entanglement of used states. On the other hand, in the
majority of real applications bright stable light beams are
required. It is therefore highly desirable to elaborate reliable
sources of light beams having the mentioned properties.

The recent development of CV quantum information is stipulated
mainly by preparation of two-mode squeezed states, which are an
approximation to EPR (Einstein-Podolsky-Rosen) entangled states
and can be easily generated by a nondegenerate parametric
amplifier \cite{Reid, Kimble}. However, up to now the generation
of bright light beams with high degree of CV entanglement meets
serious problems. One of these is the degradation of entanglement
due to uncontrolled dissipation and decoherence processes.
Nevertheless, an experimental progress in the generation of CV
entangled intensive light beams from NOPO has been reported in
\cite{Zhang}. A source of CV entangled light has been also built
using two quadrature squeezed beams combined on a beam splitter
\cite{Furusawa}, and experimental characterization of CV
entanglement has been demonstrated using this method in
\cite{Bowen}. CV entanglement of phase-locked light beams was
recently studied in \cite{PhaseLocked}.

Note that because the number of experimentally accessible and
feasible multi-wave nonlinear interactions leading to formation of
entanglement is rather limited, the class of schemes that may be
practically elaborated is restricted. In this Letter we point out
that the class of currently proposed schemes for generation of
intensive light bemas with high degree of entanglement may be
significantly extended if instead of monochromatic pumping we
consider periodically modulated pump fields. As a realization of
this idea, we propose in this Letter a novel scheme of NOPO in a
cavity, driven by an amplitude-modulated electromagnetic field and
stress that this scheme provides highly effective mechanism for
improvement of the degree of CV entanglement, even in the presence
of dissipation and cavity induced feedback.

CV entangling resources are usually analyzed based on two-mode
squeezing through the variances of the quadrature amplitudes of
two generated modes. In NOPO, the two-mode integral squeezing,
which characterizes the CV entanglement, reaches only $50\%$
relative to the level of vacuum fluctuations, if the pump field
intensity is close to the generation threshold \cite{Levon,
Dechoum}. We will demonstrate here, that the level of two-mode
squeezing in the proposed scheme is not limited, which indicates a
high degree of quadrature entanglement obeying the condition of
EPR-like paradox criterion as quantified by Reid and Drummond
\cite{Reid}.

It seems intuitively clear that such an achievement is due to the
control of quantum dissipative dynamics through the application of
suitably tailored, time-modulated driving field. Some examples of
the suppression of quantum decoherence as well as an improvement
of nonclassical statistics of oscillatory excitation numbers have
been considered in \cite{Viola}.

We consider a type-II phase-matched NOPO with triply resonant
cavity that supports the pump mode and two orthogonally polarized
modes of subharmonics. The Hamiltonian describing intracavity
interactions within the framework of rotating wave approximation
and in the interaction picture is
\begin{eqnarray}
H &=&i\hbar f\left(t\right)\left( e^{i\left( \Phi _{L}-\omega_{L}
t\right)}a_{3}^{+}-e^{-i\left( \Phi _{L}-\omega_{L} t\right)
}a_{3}\right)\nonumber \\
&&+i\hbar k\left(e^{i\Phi _{k}}a_{3}a_{1}^{+}a_{2}^{+}-e^{-i\Phi
_{k}}a_{3}^{+}a_{1}a_{2}\right),\label{Hamiltonian}
\end{eqnarray}
where $a_{i}$ are the boson operators for cavity modes at the
frequencies $\omega_{i}$. The pump mode $a_{3}$ is driven by an
amplitude-modulated external field at the frequency
$\omega_{L}=\omega_{3}$ with amplitude $f(t)$ that is a periodic
function with modulation frequency $\delta\ll\omega_{L}$. The
constant $ke^{i\Phi _{k}}$ determines an efficiency of the
down-conversion process in $\chi^{(2)}$ medium.

The system of interest is dissipative because the subharmonic
modes suffer losses. Taking into account the cavity damping rates
$\gamma _{i}$ of the modes we consider the case of zero detunings
and high cavity losses for the pump mode ($\gamma _{3}\gg \gamma$)
under the assumption that $\gamma_{1}=\gamma_{2}=\gamma$. However,
in our analysis we allow for the pump depletion effects. Following
the standard procedure of quantum optics we derive in the positive
P-repesentation \cite{Drummond} the stochastic equations for two
groups of independent complex c-number variables $\alpha_{1,2}$
and $\beta_{1,2}$ corresponding to operators $a_{1,2}$ and
$a^{+}_{1,2}$
\begin{eqnarray}
\frac{d\alpha _{1}}{dt}=-(\gamma+ \lambda \alpha_{2}\beta
_{2})\alpha_{1}+\varepsilon\left(t\right)\beta
_{2}+W_{\alpha_{1}}\left(t\right),
\label{alpha1StochEq} \\
\frac{d\beta _{1}}{dt}=-(\gamma+\lambda
\alpha_{2}\beta_{2})\beta_{1}+\varepsilon^{\ast}
\left(t\right)\alpha_{2}+W_{\beta_{1}}\left(t\right).
\label{beta1StochEq}
\end{eqnarray}
Here: $\varepsilon(t)=f(t)k/\gamma_{3}$,
$\lambda=k^{2}/\gamma_{3}$ and equations for $\alpha_{2},
\beta_{2}$ are obtained from (\ref{alpha1StochEq}),
(\ref{beta1StochEq}) by exchanging the subscripts
(1)$\leftrightarrows$(2). Our derivation is based on the Ito
stochastic calculus, and the nonzero stochastic correlations are:
$\langle W_{\alpha_{1}}\left(t\right)W_{\alpha_{2}}
\left(t^{\prime}\right)\rangle = \left(\varepsilon\left(t\right)
-\lambda \alpha _{1}\alpha
_{2}\right)\delta\left(t-t^{\prime}\right)$, $\langle
W_{\beta_{1}}\left(t\right)W_{\beta_{2}}\left(t^{\prime}\right)\rangle=\left(\varepsilon\left(t\right)
-\lambda \beta _{1}\beta
_{2}\right)\delta\left(t-t^{\prime}\right)$. Note, that while
obtaining these equations we used the transformed boson operators
$a_{i}\rightarrow a_{i}exp\left(-i\Phi_{i}\right)$ with $\Phi_{i}$
being $\Phi_{3}=\Phi_{L}$,
$\Phi_{1}=\Phi_{2}=\frac{1}{2}\left(\Phi_{L}+\Phi_{k}\right)$.
This leads to cancellation of phases at intermediate stages of
calculation. As a result, the equations depend only on real and
positive coupling constants.

The equations of motion are with time-dependent coefficient.
Nevertheless, surprisingly, it is possible to find their
analytical solution in the semiclassical approach for an
arbitrary, but real modulation amplitude $f(t)$. The presentation
of both the semiclassical and quantum theories of such a
time-modulated NOPO is another important goal of this Letter.

First, we shall study the solution of stochastic equations in
semiclassical treatment, neglecting the noise terms, for mean
photon numbers $n_{j}$ and phases $\varphi_{j}$ of the modes
($n_{j}=\alpha_{j}\beta_{j}$,
$\varphi_{j}=ln(\alpha_{j}/\beta_{j})$) for time-intervals
exceeding the transient time, $t\gg\gamma^{-1}$. An analysis shows
that similar to the standard NOPO, the considered system also
exhibits threshold behavior, which is easily described through the
period-averaged pump field amplitude
$\overline{f(t)}=\frac{1}{T}\int^{T}_{0}f(t)dt$, where
$T=2\pi/\delta$. The below-threshold regime with a stable trivial
zero-amplitude solution is realized for $\overline{f}<f_{th}$,
where $f_{th}=\gamma\gamma_{3}/k$ is the threshold value. When
$\overline{f}>f_{th}$, the stable nontrivial solution exists with
the following properties. First, the mean photon numbers for
subharmonic modes
$n_{oi}=\left<a_{i}^{+}a_{i}\right>=\left|\alpha_{i}\right|^{2}$
are equal one to the other ($n_{01}=n_{02}=n_{0}$) due to the
symmetry of the system, $\gamma_{1}=\gamma_{2}=\gamma$. As for
usual NOPO, the phase difference is undefined due to the phase
diffusion, while the sum of phases is equal
$\varphi_{1}+\varphi_{2}=2\pi k$. The mean photon number
$n_{0}(t)$ satisfies the following equation
\begin{equation}
\frac{d}{dt}n_{0}\left(t\right)=2n_{0}\left(t\right)\left(\varepsilon\left(t\right)-\gamma-\lambda
n_{0}\left(t\right)\right),\label{classicalEquationOfPhotonNumber}
\end{equation}
which can be explicitly integrated, giving a periodic solution
\begin{equation}
n_{0}^{-1}(t)=2\lambda\int^{0}_{-\infty}exp\left(2\int^{\tau}_{0}
\left(\varepsilon\left(t^{\prime}+t\right)-\gamma\right)dt^{\prime}\right)d\tau.\label{asymptoticSolution}
\end{equation}
Note, that in (\ref{asymptoticSolution}) the terms corresponding
to the transient dynamics are omitted for simplicity. In the
absence of modulation $n_{0}(t)$ reaches the steady-state
$n_{0}=(\overline{f}-f_{th})/k$ in accordance with the theory of
standard NOPO for zero detunigs.

\begin{figure}
\includegraphics[angle=-90,width=0.48\textwidth]{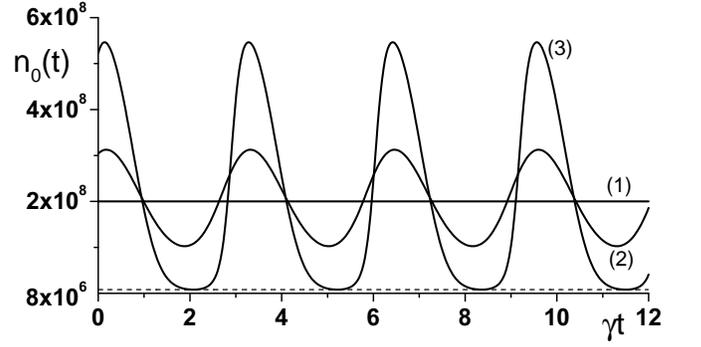}
\caption{ The semiclassical mean photon number versus
dimensionless time for the parameters: $k/\gamma=5\cdot10^{-4}$,
$\gamma_{3}/\gamma=25$, $\delta/\gamma=2$, $\overline{f}=3f_{th}$:
$f_{1}=0$ (curve 1), $f_{1}=0.4\overline{f}$ (curve 2) and
$f_{1}=1.2\overline{f}$ (curve 3).} \label{Photon_fig}
\end{figure}

We emphasize that this model is available for experiments. It can
be implemented at least for NOPO driven by polychromatic pump
fields. Particularly, for the following pump field
$E_{ext}(t)=\overline{f}\cos(\omega_{L}t+\Phi_{L})+
(f_{1}/2)(\cos((\omega_{L}+\delta)t+\Phi_{L}+\Phi)
+\cos((\omega_{L}-\delta)t+\Phi_{L}-\Phi))$ the system is
described by the Hamiltonian (\ref{Hamiltonian}) with harmonic
modulation amplitude $f(t)=\overline{f}+f_{1}\cos(\delta t+\Phi)$.
We present below the final results for the case of such harmonic
modulation assuming without loss of generality $\overline{f}>0$,
$f_{1}>0$ and $\Phi = 0$. In this case the mean photon number
(\ref{asymptoticSolution}) reads as
\begin{eqnarray}
n_{0}^{-1}(t)&=&2\lambda\int^{0}_{-\infty}\exp\left(2\gamma\tau\left(\frac{\overline{f}}{f_{th}}-1\right)\right)\times\nonumber\\
&&\exp\left(\frac{2\gamma f_{1}}{\delta
f_{th}}\left[\sin\left(\delta\left(
t+\tau\right)\right)-\sin\left(\delta t\right)\right]
\right)d\tau.\label{asymptoticSolutionForNForHarmonicModulation}
\end{eqnarray}
This result is illustrated in Fig.(\ref{Photon_fig}) for the
different levels of modulation.

To characterize the CV entanglement we address to both the
inseparability criterion proposed by Duan et al. and Simon
\cite{Simon} and the EPR paradox criterion \cite{Reid}. These
criteria could be quantified by analyzing the variances of a
relevant distance $V_{-}=V\left(X_{1}-X_{2}\right)$ and total
momentum $V_{+}=V\left(Y_{1}+Y_{2}\right)$ in the terms of the
quadrature amplitudes of two modes
$X_{k}=X_{k}\left(\Theta_{k}\right)=\frac{1}{\sqrt{2}}
\left(a^{+}_{k}e^{-i\Theta_{k}}+a_{k}e^{i\Theta_{k}}\right)$, $
 Y_{k}=X_{k}\left(\Theta_{k}-\frac{\pi}{2}\right),\;\left(k=1,2\right)$,
where $V(x)=\left<x^{2}\right>-\left<x\right>^{2}$ is a denotation
of the variance. The inseparability criterion, or weak
entanglement criterion, for the sum of variances quite generally
reads as $V_{+}+V_{-}<2$, and due to the mentioned symmetries is
reduced to the following form $V=V_{+}=V_{-}<1$, while for the
product of variances this criterion has the form
$V_{+}V_{-}=V^{2}<1$. The strong CV entanglement criterion shows
that when the inequality $V_{+}V_{-}<1/4$ is satisfied, there
arises an EPR-like paradox. Obviously, the sufficiency condition
for inseparability is weaker than the EPR condition. Both criteria
have been used to characterize the entanglement mainly in spectral
measurements \cite{Bowen}. Contrary to that, we confine ourselves
to analyzing only the total intracavity variances. These variances
are expressed through the stochastic variables using the
relationships between normally-ordered operator averages and
stochastic moments with respect to the P-function. Restoring the
previous phase structure of intracavity interaction, we obtain
that $V_{+}=V_{-}=V$ and
\begin{eqnarray}
V=1+\left<\alpha_{1}\beta_{1}\right>+\left<\alpha_{2}\beta_{2}\right>
-\left<\alpha_{1}\alpha_{2}\right>e^{i\Theta}-\left<\beta_{1}\beta_{2}\right>e^{-i\Theta},\label{VarianceWithPhases}
\end{eqnarray}
where $\Theta=\Theta_{1}+\Theta_{2}+\Phi_{L}+\Phi_{k}$.

Further, we will calculate the variance in the standard linear
treatment of quantum fluctuations. To this end, it is convenient
to use the following moments of quantum variables, which are given
below for both operator- and P-representations:
$\left<n_{+}\right>:=\left<a_{1}^{+}a_{1}\right>
+\left<a_{2}^{+}a_{2}\right>=\left<\alpha_{1}\beta_{1}\right>
+\left<\alpha_{2}\beta_{2}\right>$,
$\left<R\right>:=\left<a_{1}^{+}a_{1}\right>+\left<a_{2}^{+}a_{2}\right>
-\left<a_{1}a_{2}\right>-\left<a_{1}^{+}a_{2}^{+}\right>
=\left<\left(\alpha_{1}-\beta_{2}\right)\left(\beta_{1}-\alpha_{2}\right)\right>$,
$\left<Z\right>:=\left<\left(a_{1}^{+}a_{1}-a_{2}^{+}a_{2}\right)^{2}\right>
=\left<\left(\alpha_{1}\beta_{1}-\alpha_{2}\beta_{2}\right)^{2}\right>
+\left<\alpha_{1}\beta_{1}\right>+\left<\alpha_{2}\beta_{2}\right>$.
As can be seen, the possible minimal level of variance, realized
under appropriate selection of phases
$\Theta_{1}+\Theta_{2}=-\Phi_{L}-\Phi_{k}$ in formula
(\ref{VarianceWithPhases}), is expressed through the moment
$\left<R\right>$ as $V(t)=1+\left<R(t)\right>$. Using It\^{o}
rules for changing the stochastic variables, we obtain from
(\ref{alpha1StochEq}), (\ref{beta1StochEq}) the following
equations:
\begin{eqnarray}
\frac{d}{dt}\left<n_{+}\right>&=&\left(2\varepsilon\left(t\right)-2\gamma-\lambda\right)\left<n_{+}\right>-\lambda
\left<n_{+}^{2}\right>\nonumber\\
&-&2\varepsilon\left(t\right)\left<R\right>+\lambda
\left<Z\right>,\label{stochasticEquationForNplus}\\
\frac{d}{dt}\left<R\right>&=&-\left(2\varepsilon\left(t\right)
+2\gamma+\lambda\right)\left<R\right>-\lambda
\left<n_{+}R\right>\nonumber\\
&-&2\varepsilon\left(t\right)+\lambda
\left<Z\right>,\label{stochasticEquationForRminus}\\
\frac{d}{dt}\left<Z\right>&=&-4\gamma\left<Z\right>
+2\gamma\left<n_{+}\right>.\label{stochasticEquationForNminus2}
\end{eqnarray}
From Eq.(\ref{stochasticEquationForNminus2}) $\left<Z\right>$ can
be expressed as a function of $\left<n_{+}\right>$. Substituting
this expression into (\ref{stochasticEquationForNplus}),
(\ref{stochasticEquationForRminus}) we get the following equations
which are convenient for the perturbative analysis of quantum
fluctuations
\begin{eqnarray}
\frac{d}{dt}\left<n_{+}\right>&=&\left(2\varepsilon\left(t\right)-2\gamma-\lambda\right)\left<n_{+}\right>-\lambda
\left<n_{+}^{2}\right>-2\varepsilon\left(t\right)\left<R\right>\nonumber\\
&+&2\gamma\lambda\int_{-\infty}^{t}e^{4\gamma\left(\tau-t\right)}{\left<n_{+}\left(\tau\right)\right>d\tau},\label{stochasticEquationForNplusSubstituted}\\
\frac{d}{dt}\left<R\right>&=&-\left(2\varepsilon\left(t\right)
+2\gamma+\lambda\right)\left<R\right>-\lambda
\left<n_{+}R\right>\nonumber\\
&-&2\varepsilon\left(t\right)+2\gamma\lambda\int_{-\infty}^{t}e^{4\gamma\left(\tau-t\right)}{\left<n_{+}\left(\tau\right)\right>d\tau}.\label{stochasticEquationForRminusSubstituted}
\end{eqnarray}
First, we consider the above-threshold regime linearizing quantum
fluctuations around the stable semiclassical solutions. In the
linear treatment of quantum fluctuations we have the expansions
$\left<n_{+}\right>=n_{10}+n_{20}+\left<\delta
n_{+}\right>=2n_{0}+\left<\delta n_{+}\right>$,
$\left<R\right>=R^{0}+\left<\delta R\right>=\left<\delta
R\right>$, $\left<n_{+}R\right>=2n_{0}\left<\delta R\right>$,
$\left<n_{+}^{2}\right>=4n_{0}\left<\delta n_{+}\right>$, where it
was assumed that $n_{10}=n_{20}=n_{0}(t)$,
$\varphi_{1}+\varphi_{2}=2\pi k$, and hence $R^{0}=0$. Note, that
in the current experiments the ratio of nonlinearity to dumping is
small, $k/\gamma\ll 1$ (typically $10^{-6}$ or less), and hence
$\lambda/\gamma=k^{2}/\left(\gamma\gamma_{3}\right)\ll 1$ is the
small parameter of the theory. Therefore, the zero order terms in
the above expansion correspond to a large classical field of the
order $\gamma/\lambda$ in accordance with
Eq.(\ref{asymptoticSolution}), while the next terms describing the
quantum fluctuations are of the order of $1$. On the whole,
combining the procedure of linearization with $\lambda/\gamma\ll
1$ approximation we get a linear equation for the variance
$V(t)=1+\left<\delta R\right>$
\begin{eqnarray}
\frac{d}{dt}V\left(t\right)&=&-2\left(\gamma+\varepsilon\left(t\right)+\lambda
n_{0}\left(t\right)\right)V\left(t\right) + 2\lambda n_{0}\left(t\right)\nonumber\\
&+& 2\gamma
+4\gamma\lambda\int_{-\infty}^{t}e^{4\gamma\left(\tau-t\right)}{n_{0}(\tau)d\tau},\label{linearizedV}
\end{eqnarray}
with the following periodic asymptotic solution
\begin{eqnarray}
V\left(t\right)=2\int_{-\infty}^{t}{\exp\left(-2\int_{\tau}^{t}{\left(\gamma
+\varepsilon\left(t^{\prime}\right)+\lambda n_{0}\left(t^{\prime
}\right)\right)dt^{\prime}}\right)}\times\nonumber\\
\left[\gamma+\lambda
n_{0}\left(\tau\right)+2\gamma\lambda\int_{-\infty}^{\tau}e^{4\gamma\left(\tau^{\prime}-\tau\right)}
n_{0}(\tau^{\prime})d\tau^{\prime}\right]d\tau.
\label{asymtoticSolutionV}
\end{eqnarray}
It should be noted, that the result (\ref{asymtoticSolutionV}) is
also obtained from initial equations (\ref{alpha1StochEq}),
(\ref{beta1StochEq}) in the photon number and phase variables
using, however, more complicated calculations. In the absence of
modulation the formula (\ref{asymtoticSolutionV}) coincides with
an analogous one for ordinary NOPO. The analysis of the
below-threshold regime is more simple and leads to formula
(\ref{asymtoticSolutionV}) with $n_{0}=0$.
\begin{figure}
\includegraphics[angle=-90,width=0.48\textwidth]{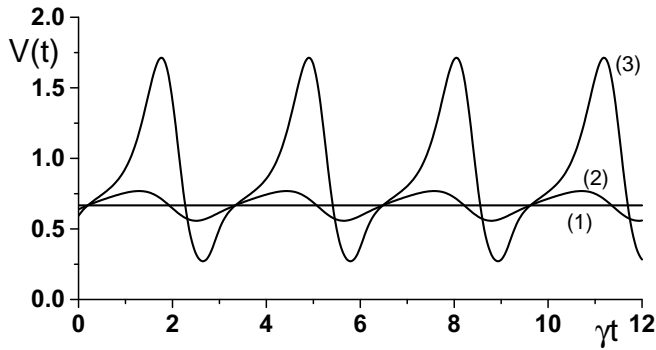}
\caption{The variance $V(t)$ given by the linear theory, versus
the dimensionless time $\gamma t$ for the parameters as in
Fig.(\ref{Photon_fig}).} \label{Variance_fig}
\end{figure}

\begin{figure}
\includegraphics[angle=-90,width=0.48\textwidth]{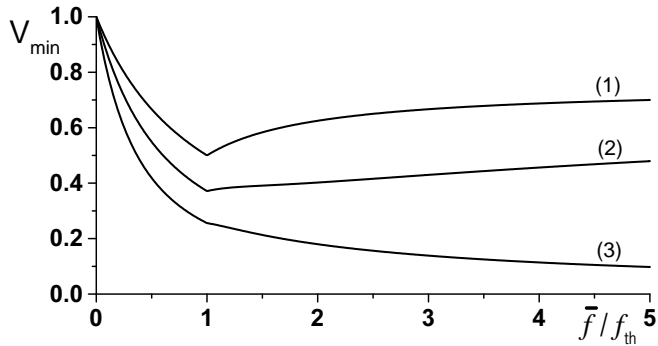}
\caption{The minimum level of the variance versus
$\overline{f}/f_{th}$ for three levels of modulation: $f_{1}=0$
(curve 1), $f_{1}=0.75\overline{f}$ (curve 2) and
$f_{1}=2\overline{f}$ (curve 3). The parameters are:
$k/\gamma=5\cdot10^{-4}$, $\gamma_{3}/\gamma=25$,
$\delta/\gamma=2$.} \label{Vmin_fig}
\end{figure}

Let us outline some conclusions from this result for the case of
harmonic modulation, using that $\varepsilon
\left(t\right)=\frac{k}{\gamma_{3}}\left(\overline{f}+f_{1}\cos\left(\delta
t\right)\right)$ and the expression
(\ref{asymptoticSolutionForNForHarmonicModulation}). Typical
results are presented in Fig.(\ref{Variance_fig}) for the
above-threshold regime, $\overline{f}/f_{th}=3$. The variance is
seen to show a time-dependent modulation with a period
$2\pi/\delta$. The drastic difference between the degree of
two-mode squeezing/entanglement for modulated and stationary
dynamics is also clearly seen in Fig.(\ref{Variance_fig}). The
stationary variance (curve 1) near the threshold having a limiting
squeezing of $0.5$ (see also Fig.(\ref{Vmin_fig}), curve 1) is
bounded by quantum inseparability criterion $V<1$, while the
variance for the case of modulated dynamics obeys the EPR
criterion $V^{2}<1/4$ of strong CV entanglement for definite time
intervals. In particular, the minimum values of the variance and
corresponding photon numbers of Fig.(\ref{Photon_fig}) at fixed
time intervals $t_{m}=t_{0}+2\pi m/\delta$, ($m=0,1,2...$) are:
$n_{0}\simeq 6.16\cdot10^{7}$, $V_{min}\simeq 0.27$,
$t_{0}=2.64\gamma^{-1}$ (curve 3) and $n_{0}\simeq
1.71\cdot10^{8}$, $V_{min}\simeq 0.56$, $t_{0}=2.51\gamma^{-1}$
(curve 2). The dependence of $V_{min}$ on the period-averaged pump
field amplitude is shown in Fig.(\ref{Vmin_fig}) for different
levels of modulation. As it is expected, the degree of EPR
entanglement increases with ratio $f_{1}/\overline{f}$. Another
peculiarity here is that the stationary variance (curve 1) has a
characteristic threshold behavior (see, for example \cite{Levon})
that disappears in case of strong modulation (curve 3). We
conclude, that here nothing of the kind of principle is valid that
prohibits the reaching of approximately perfect
squeezing/entanglement for the case of strong modulation. The
production of strong entanglement occurs for the period of
modulation comparable with the characteristic time of dissipation,
$\delta\approx\gamma$. For both asymptotic cases of slow
($\delta\ll\gamma$) and fast ($\delta\gg\gamma$) modulations this
effect disappears.

\begin{figure}
\includegraphics[angle=-90,width=0.48\textwidth]{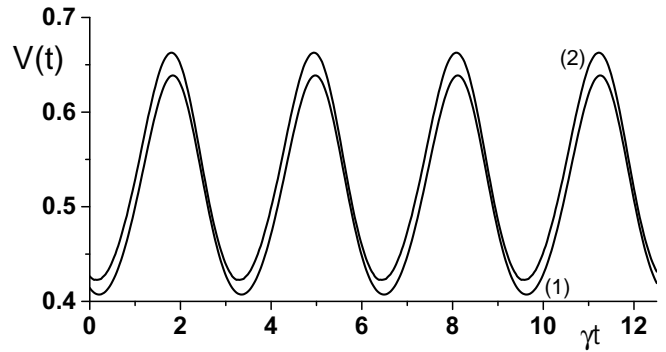}
\caption{Comparison of the squeezed variance between the
analytical solution (curve 1) and numerical simulation (curve 2).
The parameters are: $\overline{f}=f_{th}$,
$f_{1}=0.5\overline{f}$, $\lambda/\gamma=0.01$,
$\gamma_{3}/\gamma=25$, $\delta/\gamma = 2$.}
\label{Comparison_fig}
\end{figure}

The linearized theory is applicable only outside the critical
region. As our analysis shows, the condition of the validity of
linear results for the near-threshold regimes reads as
$\left|\overline{f}/f_{th}-1\right|\gg\left(\lambda/\gamma\right)
\exp\left[2(f_{1}/f_{th})(\gamma/\delta)\right]$. For typical
$\lambda/\gamma\ll 1$, this condition is fairly easy to satisfy
even for narrow critical ranges provided that
$\delta\approx\gamma$. As a rule, the quantum corrections diverge
at the classical threshold, although, the variance
(\ref{asymtoticSolutionV}) surprisingly is well defined also at
the threshold. Nevertheless, in order to verify the accuracy of
our analytical calculations we performed numerical simulation
based on the quantum state diffusion method (\cite{Gisin}, for the
case of time-modulated fields, see \cite{SRAndChaos}). This
simulation tends to disagree with our analytical results for
comparatively high values of $\lambda/\gamma$ parameter,
especially at the threshold, as is shown in
Fig.(\ref{Comparison_fig}). Generally speaking the divergence
between the numerical and analytical results is typically in the
order of $\lambda/\gamma$.

In conclusion, we have proposed a new approach to generation of
strongly EPR entangled states of intense light beams based on the
time-modulation of quantum dissipative dynamics. We have
demonstrated an essential improvement of the degree of
entanglement in modulated NOPO in comparison with the ordinary
one, if the frequency of modulation is close to the decay rate of
dissipative precesses. We have shown that surprisingly the CV
entanglement becomes approximately perfect for high-level of
modulation, but is only realized for a periodic sequence of time
intervals synchronized with the modulation period. The
analytically predicted results were confirmed in our numerical
simulation. In our study we have not analyzed in detail any
physical mechanisms leading to strong EPR entanglement due to the
modulation. This topic is currently being explored and will be the
subject of forthcoming work. We believe that the results obtained
are applicable to a general class of quantum dissipative systems
and can serve as a guide for further theoretical and experimental
studies of intensive light beams with high degree of entanglement.

\begin{acknowledgments}
Acknowledgments: This work was supported by the NFSAT PH 098-02
grant no. 12052 and ISTC grant no. A-823.
\end{acknowledgments}

\end{document}